\documentclass[a4,aps,showpacs,amsmath,floatfix]{revtex4}
\usepackage{graphicx}
\usepackage{dcolumn}
\usepackage{color}
\usepackage{amsmath}
\usepackage{here}
\begin{document}
\title{Solitons with rings and vortex rings on solitons in nonlocal nonlinear media}

\author{V.M. Biloshytskyi$^{1}$,  A.O. Oliynyk$^{1}$,  P.M. Kruglenko$^2$, A.S. Desyatnikov$^{3,4}$, A.I. Yakimenko$^{1}$}

\affiliation{$^1$  Department of Physics, Taras Shevchenko National University of Kyiv, 64/13, Volodymyrska Street, Kyiv 01601, Ukraine \\
$^2$  V. Lashkaryov Institute of Semiconductor Physics, 41 pr. Nauki, Kyiv 03028,  Ukraine
\\  $^3$ Department of Physics, School of Science and Technology, Nazarbayev University, 53 Kabanbay Batyr Ave., Astana 010000, Kazakhstan \\
$^4$ Nonlinear Physics Centre, Research School of Physics and Engineering,
The Australian National University, Canberrra, ACT 2601, Australia}

\begin{abstract}
  Nonlocality is a key feature of many physical systems since it  prevents a catastrophic collapse  and a symmetry-breaking azimuthal  instability of intense wave beams in a bulk self-focusing nonlinear media. This  opens up an intriguing perspective for stabilization of complex topological structures such as higher-order solitons, vortex rings and vortex ring-on-line complexes. Using direct numerical simulations, we find a class of  cylindrically-symmetric $n$-th order spatial solitons having the intensity distribution with a central bright spot surrounded by $n$ bright rings of varying size. We investigate dynamical properties of these higher-order solitons in a media with thermal nonlocal nonlinear response.  We show theoretically that  a vortex complex of vortex ring and vortex line,  carrying two independent winding numbers, can be created by perturbation of the stable optical vortex soliton in nonlocal nonlinear media.
\end{abstract}

\pacs{42.65.Tg, 42.65.Sf, 42.70.Df, 52.38.Hb} \maketitle

\section{Introduction}
Solitons are fundamental objects which appear in many branches of physics such as plasma, nonlinear optics, Bose-Einstein condensates (BECs), liquid crystals classical and quantum field theory, etc.
A bright spatial soliton is a wave beam of finite cross section which propagates in a
nonlinear medium without changing its structure. Spatial solitons appear as a result of a balance between diffraction  and nonlinear changes of the refractive index of the medium induced by the wave itself.
 In the spatially nonlocal media the nonlinear response depends on the wave packet intensity
at some extensive spatial domain. Nonlocality naturally arises in many nonlinear media.  In particular, a nonlocal response is
induced by heating and ionization, and it is known to be important
in media with thermal nonlinearities such as thermal
glass~\cite{thermal} and plasmas~\cite{LitvakSJPlPhys75}. Nonlocal
response is a key feature of the orientational nonlinearities due
to long-range molecular interactions in nematic liquid
crystals~\cite{ContiPRL03}. An interatomic interaction potential
in BECs with dipole-dipole interactions is
also known to be substantially nonlocal~\cite{BEC}. In all such
systems, nonlocal nonlinearity can be responsible for many novel
features such as the familiar effect of the collapse arrest
\cite{TuritsynTMF85,KrolikovskiJOptB04} and stabilization of various coherent structures.

The higher-bound solitons with field nodes (zero crossing)
have been first discovered in Ref. \cite{Yankauskas} for the local Kerr-type
nonlinear media. The $n$th bound state has a central bright
spot surrounded by $n$ bright rings of varying size. In the local nonlinear
media the higher-order solitons with zero angular momentum
show the azimuthal instability \cite{KolokolovSykov,Soto-CrespoPRA91} similar to the
instability of the vortex solitons. The rings which surround
the central peak possess a symmetry-breaking instability. As
a result, the higher-bound structures decay into several fundamental
solitons. The stationary nonspinning higher-order
solitons in the nonlocal medium with Gaussian-type response function have been investigated in Ref. \cite{PRE06} by approximate variational method.
 In Ref. \cite{PhysRevLett.98.053901}  were found analytically and numerically broad classes of  higher-order solitons representing generalization of the well-known Laguerre-Gaussian and Hermite-Gaussian linear modes for the similar model of nonlocality. The specific feature of the media with Gaussian-type nonlocal response function is that  the higher-order solitons exhibit dynamics with revival when their transformations are manifested as periodic robust oscillations between two or more spatially localized states with distinctly different symmetries  \cite{PRE06, PhysRevLett.98.053901}.
 To the best our knowledge no numerical higher-order  soliton solution was obtained so far for a nonlinear media with thermal nonlocal response function. Here we find higher-order solitons by numerical solution of the stationary nonlinear Schr\"{o}dinger equation (NLSE) with thermal optical nonlinearity. In this work we investigate stability of the higher-order  soliton by numerical simulation of the dynamical NLSE.

Vortex rings are topological structures  with a closed-loop core, which  play a crucial role in the decay of superflow and in quantum turbulence in condensed matter physics. Recently it was shown \cite{SciRep12} that in self-saturating optical nonlinear media  as a radially perturbed soliton propagates, vortex loops occur
in the form of rings perpendicular to the propagation direction. This spontaneous vortex nucleation is a consequence
of the nonlinear phase accumulation between the soliton's peak and its tail. As was discovered in Ref. \cite{SciRep12} phase singularities nucleate
if this phase difference reaches the value of $\pi$ during evolution along the optical axis $z$ in an optical media with local saturable nonlinearity. Optical vortex rings (in contrast to vortex rings in fluids) are static in time and appear when nonlinear phase of the self-trapped light beam breaks the wave front into a sequence of optical vortex loops around the perturbed fundamental soliton ($m=0$). In this paper we demonstrate that vortex rings can be generated in nonlocal nonlinear media.

A vortex line is the singular wave beam with ringlike intensity distribution, with the dark hole at the center where the phase dislocation takes place: a phase circulation around the axis of propagation is equal to $2\pi m$. In contrast to fundamental soliton ($m=0$), in a self-focusing nonlinear media the spinning solitons usually possess a strong azimuthal modulational instability. However different kinds of nonlocality of the nonlinear response can suppress or completely eliminate the symmetry-breaking azimuthal instability \cite{PRE05, PRE06, Briedis}.  This opens
up a perspective for stabilization also  vortex soliton states with complex structure such as Hopfions. A Hopfion (or Hopf soliton) is a topological soliton with two independent winding numbers: the first, $S$, characterizes a horizontal circular vortex embedded into a three-dimensional soliton; and the second, $m$, corresponds to vorticity around the axis, perpendicular
to this circle. Hopf solitons appear in many fields, including field theory, optics, ferromagnets, and semi- and superconductors.  Multicharged ($m > 1$, $S > 1$) vortex structures
have been demonstrated \cite{Berry01} to be unstable in optical
media.  Very recently stable Hopfions have been theoretically predicted in BEC \cite{Kartashov14,PhysRevA.92.053603}. In this work we demonstrate that radially perturbed vortex line soliton ($m\ge 1$) gains additional single-charged ($S=1$) vortex rings. These vortex complexes represent the first examples of optical analog of Hopfions: vortex ring-on-line.


\section{Higher-order solitons}
We consider here  nonspining ($m=0$) higher-order solitons with $n$ nodes in a nonlocal media with thermal nonlinearity.
The basic dimensionless equations describing the propagation
of the electric field envelope $\Psi(x,y,z)$ coupled to the
temperature perturbation $\theta(x,y,z)$ has the following form \cite{PRE05}:
\begin{equation}
   \begin{array}{l} {\displaystyle
       i\frac{\partial \Psi}{\partial z}+\Delta_\perp\Psi +\theta\Psi=0,
       } \\*[9pt] {\displaystyle
\alpha^2\,\theta-\Delta_\perp\theta=|\Psi|^2.
   }\end{array}
   \label{eq:NLS}
\end{equation}
Equations (\ref{eq:NLS}) describe the light propagation
in media with thermal nonlinearities, and it appears also
in the study of two-dimensional bright solitons in nematic
liquid crystals \cite{ContiPRL03} and in partially
ionized plasmas \cite{thermal}.

In the limit $\alpha^2\gg 1$, we can neglect the second term in
the equation for the field $\theta$ of Eq.~(\ref{eq:NLS}) and
reduce this system to the standard local nonlinear Schr\"{o}dinger
(NLS) equation with cubic nonlinearity. The opposite case, i.e.
$\alpha^2 \ll 1$, will be referred to as a strongly nonlocal
regime of the beam propagation.

\begin{figure*}[h]
\includegraphics[width=3.4in]{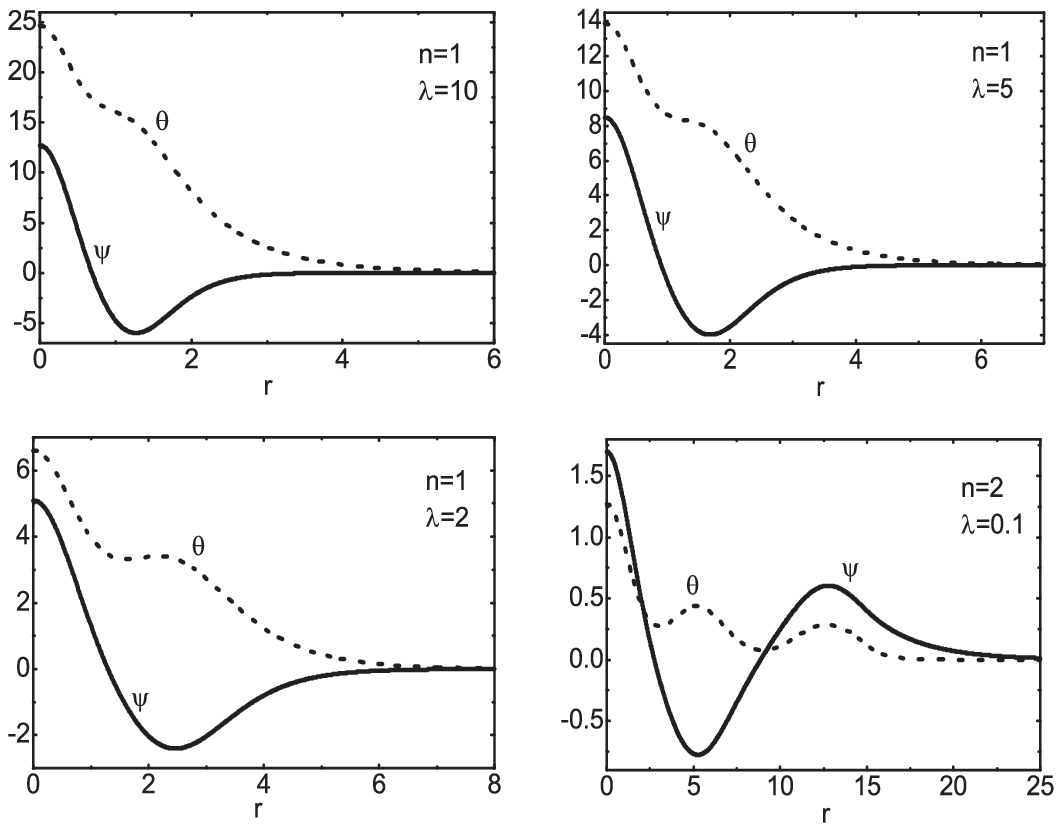}
\includegraphics[width=3.4in]{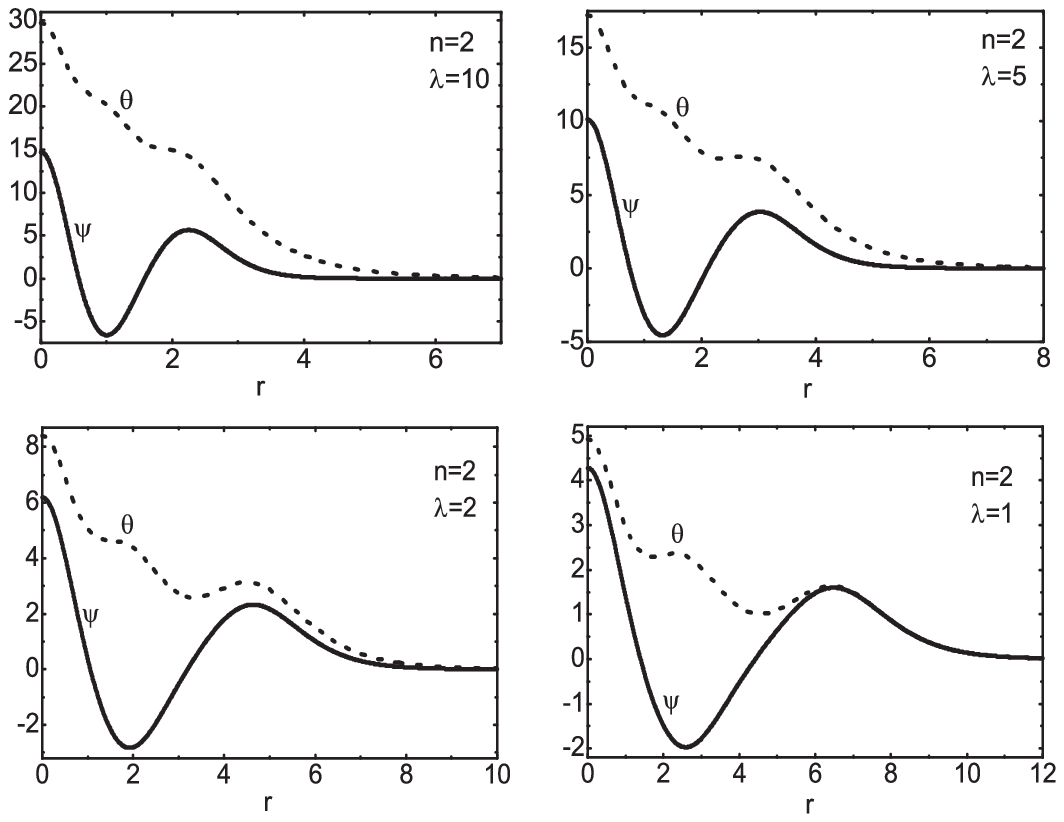}
\caption{Typical examples of numerical stationary solutions in the form of single-ring ($n=1$) and double-ring ($n=2$) optical solitons.} \label{n1cuts}
\end{figure*}



We look for the stationary solutions of the system (\ref{eq:NLS})
in the form $\Psi(x,y,z)=\psi_n(r) \exp(i\Lambda z)$,
where $r=\sqrt{x^2+y^2}$ is
the radial coordinate,  and $\Lambda$ is the beam
propagation constant. Such solutions describe either the
fundamental optical soliton, when $n=0$,  or the higher-order soliton
with  $n$ nodes, when $n > 0$.

\begin{figure}
\includegraphics[width=\textwidth]{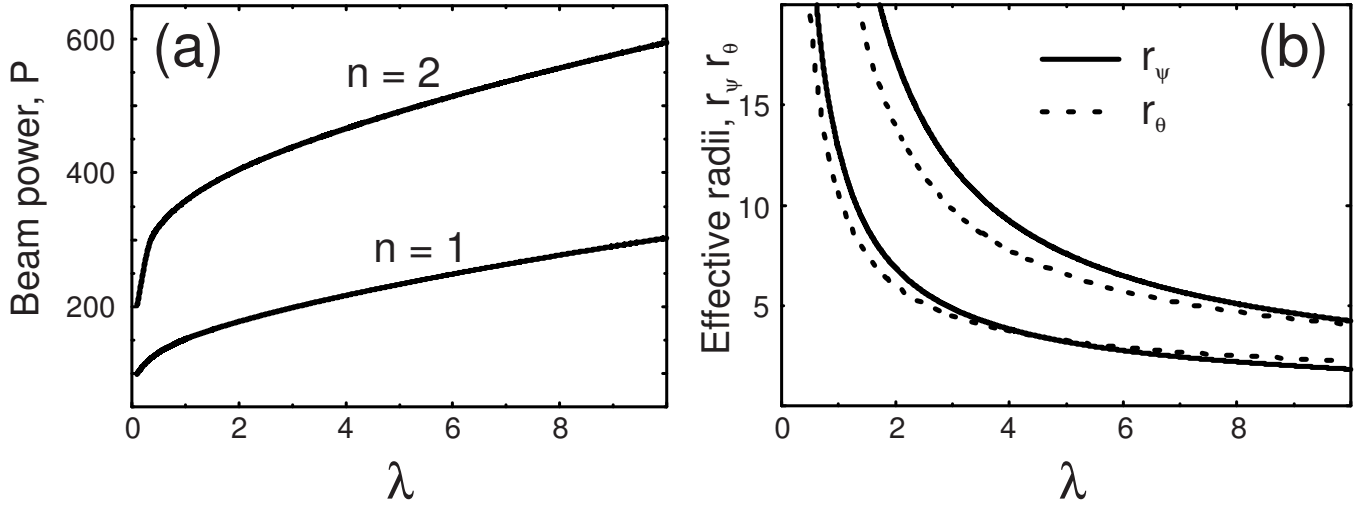}
\caption{(a) Beam power vs rescaled propagation constant $\lambda=\Lambda/\alpha^2$ for solitons with nodes ($n=1,2$).  (b) Effective radii $r_\psi$ (solid curves) and $r_\theta$ (dashed curves) for single- and double ring solitons ($n=1,2$).} \label{PandRs}
\end{figure}

\begin{figure*}[h]
\includegraphics[width=\textwidth]{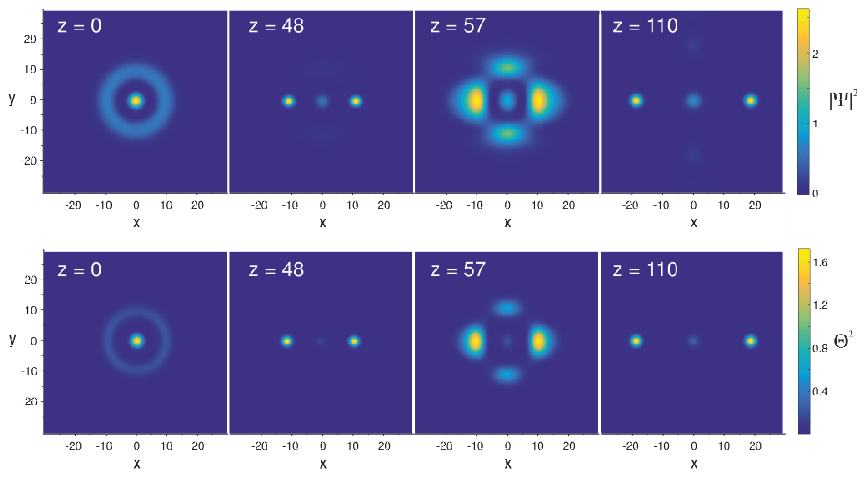}
\caption{Snapshots of the evolution in $z$-direction single-ring ($n=1$) optical solitons for $\lambda=0.1$. Shown are intensity distribution $|\Psi(x,y)|^2$ and $\Theta^2(x,y)$.} \label{Dynamics1}
\end{figure*}
\begin{figure*}[h]
\includegraphics[width=\textwidth]{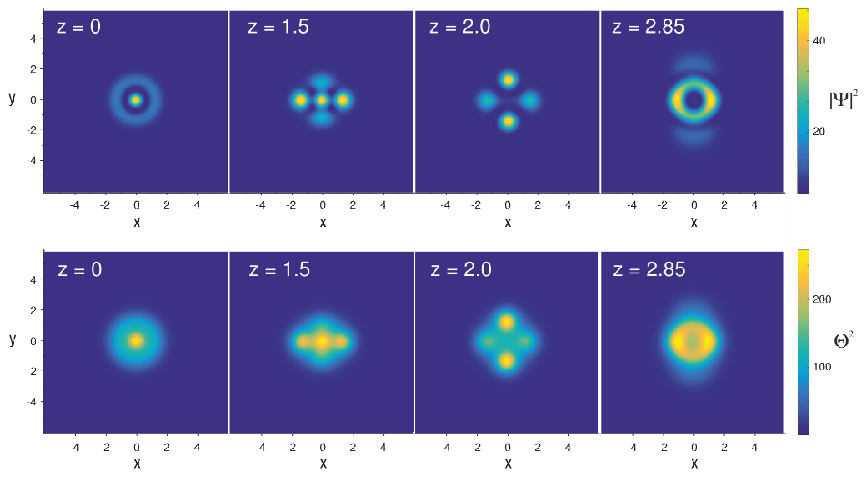}
\caption{Snapshots of the evolution in $z$-direction single-ring ($n=1$) optical solitons for $\lambda=10$. Shown are intensity distribution $|\Psi(x,y)|^2$ and $\Theta^2(x,y)$.}
\label{Dynamics2}
\end{figure*}

The beam radial profile $\psi(r)$ and the temperature field
$\theta(r)$ associated with it are be found by solving the system
of ordinary differential equations,
\begin{equation}
-\lambda\psi+\Delta_r\psi+\theta\,\psi=0;\,\,\,
\theta-\Delta_r\theta=|\psi|^2,   \label{eq:psi_r}
\end{equation}
where $\Delta_r=d^2/d r^2+(1/r)(d/d r)$, and
$\psi$, $\theta$, $1/r^2$, $\lambda=\Lambda/\alpha^2$ are rescaled by the factor
the parameter $\alpha^2$  which itself becomes one.
Boundary conditions are: for the localized field,
$\psi(\infty) =\psi'(0)=0$, and for the temperature field,
$d\theta/dr|_{r=0}=0$ and $\theta(\infty) =0$.
The system of equations (\ref{eq:psi_r}) was solved by shooting method. 
Typical examples of numerical solutions in the form of higher-order solitons with $n=1$ and $n=2$ are shown in Figs. \ref{n1cuts}.
Figure \ref{PandRs} (a) shows the beam power
 $P = \int\left|\psi \right|^2 d^2 \textbf{r}$,
 as a
function of the rescaled propagation constant $\lambda=\Lambda/\alpha^2$. Note that in strongly nonlocal regime ($\alpha^2\ll 1$, i.e. $\lambda\gg 1$) the profile of the temperature distribution is wider than the effective radius of the central bright core of the solitonic wave beam.
Let us define the effective radii $r_\psi$ and $r_\theta$ of the
intensity distribution $|\psi|^2$ and the temperature distribution $\theta$,
respectively, as follows:
$$r^2_\psi=\frac1P\int r^2|\psi(r)|^2d^2 \textbf{r}, \;\;\;
 r^2_\theta=\frac{\int r^2\theta(r)d^2 \textbf{r}}{\int\theta(r)d^2
\textbf{r}}.
$$
 Figure \ref{PandRs} (b) shows the radii  $r_\psi$ and $r_\theta$
as functions of the $\lambda$. Both  $r_\psi$ and $r_\theta$
decrease monotonically when $\lambda$
grows.

While the fundamental solitons are known to be stable in a collapse-free self-focusing nonlinear media, stability of the higher-order structures (such as vortex solitons, bound states of solitons, and other higher-order structures) crucially depends on the specific form of the nonlocality. It is remarkable that for the model with Gaussian-type response function approximate single-charge ($m=1$) vortex soliton solution with one additional ring ($n=1$) has been investigated in Ref. \cite{Briedis}. It was found that the robust propagation  has been observed in dynamical simulations initiated by approximate field envelope in the form of Laguerre-Gaussian LG$^1_1$ wave beam.

We have performed numerical simulations of dynamics of perturbed
stationary solutions. Dynamical system (\ref{eq:NLS}) was solved numerically by employing the split-step Fourier method.
The typical examples of evolution of $n=1$ soliton are shown in Fig. \ref{Dynamics1} and Fig. \ref{Dynamics2}. The higher-order solitons exhibit symmetry-breaking azimuthal instability, which leads to decay of the solitons with nodes in weakly nonlocal regime as is seen from Fig. \ref{Dynamics1}.
In our numerical simulations in strongly nonlocal regime we observed dynamics with partial revival of the nonspinning  higher-order soliton similar to observed in Ref. \cite{PRE06} for the model based on Gaussian-type kernel of the nonlocal media response function. However, in a sharp contrast to the model with the Gaussian-type response function, the higher-order solitons do not  exhibit robust oscillations between eigen states of different symmetry in a media with thermal nonlocal nonlinearity. As is seen from Fig. \ref{Dynamics2} higher-order solitons firs transform into Hermit-Gauss mode but than rapidly decays. Solitons with more rings ( $n\ge 2$) decay in a similar way even faster than one-node ($n=1$) soliton.

\section{Vortex rings and vortex ring-on-line}

A vortex ring is one of the most universal wave structures in fields of different nature. Vortex rings have been the subject of numerous studies from classical fluid mechanics \cite{Saffman} to optics \cite{SciRep12}. It is of special interest to construct stable vortex soliton
states with complex structures, such as vortex knots, three-dimensional Skyrmions, and vortex ring-on-line.

\begin{figure*}[h]
\includegraphics[width=\textwidth]{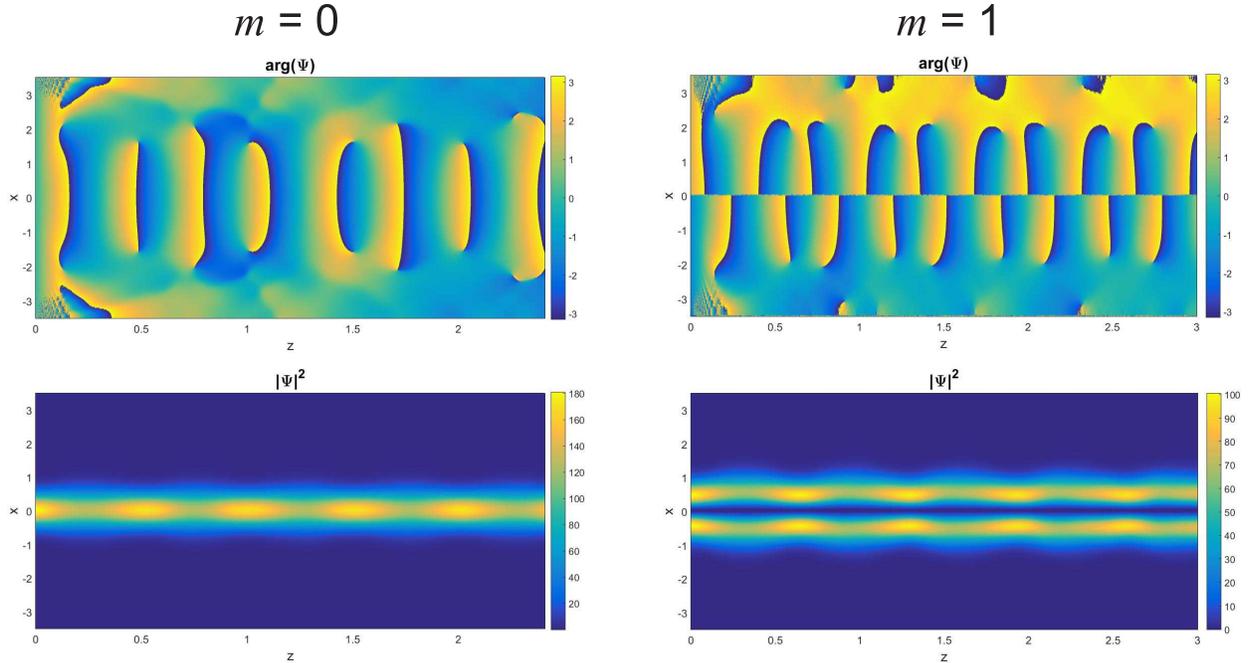}
\caption{Phase and intensity distributions for initially ''stretched'' (with $a=b=1.1$)  soliton (left) and single-charged vortex (right).  Vortex rings are clearly seen as the phase dislocations in the longitudinal plane $(x,z)$ for $\lambda=\Lambda/\alpha^2=15$.} \label{Fig_Hopfions}
\end{figure*}


In our recent works  we suggested an experimentally
feasible trapping configuration that can be used to create,
stabilize, and manipulate a vortex ring in a controllable
and nondestructive manner \cite{PRA13}; using a similar
trapping potential and a rotating condensate, we demonstrated both
energetic and dynamical stability of the Hopf soliton in BEC \cite{Hopfion}.

Here we investigate vortex rings and vortex ring-on-line in nonlocal nonlinear media with Guassian-type response function. As known, in this model not only single-charge ($m=1$) but also multiply-charged vortices are stable \cite{PRE06}. The envelope of the electric field $\Psi(x,y,z)$ obeys the NLSE:
\begin{equation}\label{eq:NLSgauss}
i\frac{\partial \Psi}{\partial z}+\Delta_\perp\Psi +\Theta \Psi=0,
\end{equation}
where
\begin{equation}  \label{eq:Theta}
\Theta(\vec r)=\int{R(|\vec{r}-\vec{r_1}|)
|\Psi(\vec{r_1})|}^2d^2\textbf{r}_1.
\end{equation}
\begin{equation}
\label{kernel}
 R(|\vec{r}-\vec{r_1}|)=\frac{\alpha^2}{\pi}
e^{-\alpha^2\left|\vec{r}-\vec{r_1}\right|^2},
\end{equation}
 $\alpha$ is the nonlocality parameter.

 Vortex solitons are the stationary solutions of the form
 \begin{equation}\label{eq:psistationary}
\Psi(x,y,z)=\psi(r) e^{im\varphi+i\Lambda z},
\end{equation}
 where $m$ is the topological charge, $\Lambda$ -- is the propagation constant, $\Psi$ satisfies the stationary NLSE:
  \begin{equation}\label{eq:psi_Vs_r}
-\Lambda\psi+\Delta_r^{(m)}\psi+\theta\,\psi=0,
\end{equation}
where
\begin{equation}\label{eq:thetaSteady}
 \theta(r)=2\alpha^2\int_0^{+\infty}e^{-\alpha^2(r-r_1)^2}\mathcal{I}_0(2\alpha^2 r r_1)|\psi(r_1)|^2 r_1d
 r_1,
\end{equation}
and $\mathcal{I}_\nu(x)=e^{-x}I_\nu(x)$, $I_\nu(x)$ -- is the modified Bessel function. In Ref. \cite{PRE06} we have investigated stationary soliton and vortex solutions.
Here we simulate evolution along the optical axis $z$ of the soliton ($m=0$) and single-charge vortex ($m=1$) with initial condition of the form:
$$
\Psi(x,y, z=0)=\psi\left(\sqrt{x^2/a^2+y^2/b^2}\right)e^{im\varphi}
$$
The initial ''stretching'' with $a=b \ne 1$ leads to radial oscillations and additional nonlinear accumulation of phase.
Left column in Fig. \ref{Fig_Hopfions}  demonstrates this process for perturbed fundamental soliton ($m=0$) with the field's topological structure, namely the
appearance of a regular set of vortex rings similar to vortex rings revealed in \cite{SciRep12} for  media with local saturating nonlinearity. The right column in Fig.  \ref{Fig_Hopfions} shows the previously
unknown features of the field's topological structure, namely the
appearance of a regular set of  vortex ring-on-line structure.





\section{Summary and conclusions}
We investigate ring-shaped  solitonic and vortex complexes in nonlocal nonlinear media.  Using direct numerical simulations, we find the $n$th bound solitonic state which has a central bright spot surrounded by $n$ rings of varying size.  These structures decay into several fundamental solitons  due to symmetry-breaking azimuthal instability in a weakly nonlocal regime, but they exhibit dynamics with partial revivals for highly-nonlocal regime.
 We show that  vortex rings appear at the periphery of the perturbed fundamental soliton while the  vortex ring-on-line complex can be created by perturbation of the stable vortex soliton. Our results open up an intriguing perspective for observation of novel complex vortex structures in nonlocal optical media, nematics, and plasmas.

\section*{ACKNOWLEDGMENTS}
A.Y., V.B., and A.O. acknowledge support from Project 1/30-2015 "Dynamics
and topological structures in Bose-Einstein condensates of
ultracold gases" of the KNU Branch Target Training at the
NAS of Ukraine.


\end{document}